\documentclass{article}
\usepackage[T1]{fontenc}
\usepackage{cite}
\usepackage{amsmath,amssymb,amsfonts}
\usepackage{algorithmic}
\usepackage[]{graphicx}
\usepackage[mathlines,switch]{lineno}
\usepackage{multicol,multirow}
\usepackage{epsfig}
\usepackage{float} 
\usepackage[left=1in,right=1in,top=1in,bottom=1in]{geometry}
\usepackage{textcomp}
\usepackage{wrapfig,colortbl}

\title{The Magnetic-Resonance-Based, Omni-Directional Hydrophone}
\author{Nathan Thyberg \thanks{Electrical and Computer Engineering, Brigham Young University, Provo, Utah, USA.} \thanks{Corresponding Author: (e-mail: natha912@student.byu.edu).} \thanks{``This work was supported in part by the National Science Foundation under Grant 2138403, the National Institutes of Health under grant R01EB032773 and the Focused Ultrasound Foundation Global Intern Program.'' }\and Davi Cavinatto \thanks{Department of Radiology, University of Utah, Salt Lake City, UT, USA.}\and
Katia Oler \footnotemark[1]\and
Elijah Oxborrow \footnotemark[1]\and
Hyrum Mangum \footnotemark[1]\and
Steven P. Allen \footnotemark[1] \footnotemark[3]}
\date{}

\begin{document}

\maketitle

\begin{figure}[h]
\centerline{\includegraphics[width=300pt]{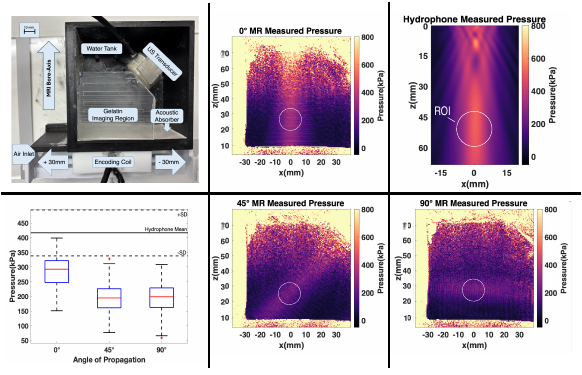}}
\label{fig:Abstract}
\end{figure}

\begin{abstract}
This study aims to improve the directional sensitivity of magnetic resonance hydrophone (MRH) methods and allow for detection and quantification of waves traveling at any angle relative to the MRI bore-axis. Doing so may better integrate MRH technology into transcranial focused ultrasound neuromodulation therapies. Existing MRH implementations only encode acoustic waves propagating along the MRI bore-axis, precluding use with wide aperture transducers or temporal skull windows. This study expands the MRH technique into a MR-based Omni-directional Hydrophone (MRbOH) that can quantify the magnitude pressure and phase of waves propagating at any angle. The MRbOH method formulates a system of equations around uniquely sensitized images that, when solved, introduce sensitivity to oblique acoustic waves. The method was tested experimentally in a gel phantom that was insonated at $0^o$, $45^o$, and $90^o$ relative to the MRI bore-axis. The method reconstructed pressure maps at all three angles of propagation, including features such as peak pressure, partial reflections, and beamwidth. The MRbOH method measured an average beamwidth of 16.2 mm while the hydrophone measured a beamwidth of 14.5 mm. Average hydrophone-measured pressure within a region of interest was 416.19 $\boldsymbol{\pm}$ 78.07 kPa. Average MRbOH measured pressure in the same region at $0^o$, $45^o$, and $90^o$, respectively, were: 285.5 $\boldsymbol{\pm}$ 49.5 kPa, 193.13 $\boldsymbol{\pm}$ 43.55 kPa, 194.7 $\boldsymbol{\pm}$ 47.38 kPa. In addition, the study presents and experimentally verifies a statistical noise model that predicted the mean with $R^2 = 0.85$, and standard deviation with $R^2 = 0.83$.
\end{abstract}
\vspace{1\baselineskip}\vspace*{-1pt}





\section{Introduction}

Transcranial ultrasound neuromodulation (TUSN) promises to treat mental health disorders by transmitting low intensity ultrasound waves through the intact scalp and skull into the brain. The thermal and mechanical energy  within the acoustic field then modulate neural activity~\cite{Tufail2011,Legon2018,Legon2018a}. Unlike their electromagnetic counterparts, acoustic fields can be shaped to produce foci at nearly any depth within the patient, resulting in flexible and focal treatment throughout the human brain~\cite{Clement2002a, Aubry2003_expDemonstration} . TUSN promises to combine non-invasive-ness, depth penetration, focal effects, and flexible steering to modulate neural networks.
Skull-based attenuation and aberration presents a persistent barrier to reliable TUSN. Human skulls possess incredible diversity in acoustic properties ~\cite{Pinton2012_attenuation},~\cite{Pichardo2011_acousticPropsHumanSkulls} that can conspire to attenuate and abberate acoustic fields inside the brain. Additionally, these effects change with location and angle of ultrasound incidence~\cite{riis2022_Properties,Ammi2008CharacterizationBone.}, the composition of each layer of the skull~\cite{Fry1978AcousticalSkull., White1978TheSkull.}, and even the distance between the skull surface and the transducer~\cite{Li2025_skullProps}. Without patient-specific compensation for these alterations, the amplitude, shape, and location of the acoustic focus can deviate from patient to patient. This large uncertainty in delivered acoustic energy at target can translate into uncertainty in treatment outcomes.

Existing TUSN clinical trials and case reports address patient specific skull variation via two methods. First, clinicians can unilaterally limit sonication intensity by assuming that all patients possess transmissive skulls~\cite{Mahoney2023,Strohman2024,Oh2024_depression}. Accordingly, only those subjects with inadvertently favorable skulls receive the intended acoustic intensity---resulting in positive case reports but ambiguous clinical trial results. 
Second, clinicians can utilize computational predictions of acoustic intensity at target~\cite{Aubry2003_expDemonstration,Pinton2012_attenuation,Pichardo2011_acousticPropsHumanSkulls,Webb2022,Webb2018,Webb2021_atten}. These methods map patient-specific features obtained with medical imaging to skull acoustic properties~\cite{Clement2002a, McDannold2019_attenuationFromCT}. They then simulate acoustic propagation through the skull and into the target. Computational methods do reduce uncertainty in the acoustic intensity at target and successfully predict the location and shape of the acoustic field~\cite{Leung2019_rapid}.  However, they still fail to predict variations in the delivered intensity across patients \cite{Leung2019_rapid, Krokhmal2025_kWave}. Even with extensive clinical experience, device manufacturers can guarantee no better certainty in delivered pressure than $\pm$40\%~\cite{Klein-Flugge2025_OpenLetter}.

In pursuit of reduced uncertainty in delivered pressure during TUSN, our team previously developed a human-compatible magnetic resonance hydrophone (MRH) prototype~\cite{MRH2025} that can non-invasively visualize acoustic wave propagation in tissue mimicking gels. The MRH encodes the correlation between acoustic displacement and a sensitizing magnetic field gradient into the voxel phase of MR images at locations and depths that correspond to the brain cortex. Under a plane wave approximation, acoustic pressure at each voxel location can then be calculated from MR voxel phase. The MRH promises to directly measure skull abberation. In a gel phantom, and using a single element, 500 kHz transducer coupled to a wave guide, the MRH produced pressure quantification error on the order of 10\%, and, on average, underestimated pressures measured by a needle hydrophone by 12 kPa~\cite{MRH2025}. Preliminary MRH data have also demonstrated encoding through an ex-vivo skull flap ~\cite{Cavinatto2025}.

While promising, current MRH designs suffer from constraints that limit clinical applicability. Published MRH studies sense acoustic waves propagating along the axis of the MRI scanner bore\cite{Plewes2000,Walker1998}, which corresponds to the cranial-caudal axis under conventional patient positioning. However, the temporal and occipital bones present more favorable acoustic windows for many TUSN targets. Current MRH designs are insensitive to acoustic propagation through these bones. Additionally, practical TUSN beams are focused and will always contain components orthogonal to the MRH's sensitive axis, regardless of the transducer's orientation.

To address these constraints, we hypothesize that the MRH technique can be expanded into an MR-based omnidirectional hydrophone (MRbOH) that can detect and quantify acoustic propagation along any arbitrary axis. This study, building on our preliminary results ~\cite{Thyberg2025},~\cite{Thyberg2026_6d0276cd},~\cite{Thyberg2026}, describes the theoretical underpinnings of this expansion and examines its performance in a tissue mimicking gel phantom.

\section{Theory} \label{sec:Theory}
The goal of this section is to develop a mathematical and statistical noise model of the MRbOH technique and identify metrics against which optimal MRbOH features can be designed.

\subsection{Signal Model} \label{sec:theory}
We develop the MR-based omnidirectional hydrophone by expanding, for a monochromatic plane wave, the relation between MR voxel phase, $ \theta$, and maximum acoustic displacement, $\xi$ , first identified by Walker and Plewes~\cite{Walker1998}. We allow the plane wave to produce displacement according to arbitrary vector $\vec{\xi}$ and undergo MRH encoding by an oscillating magnetic field gradient  $\vec{G}(t)$ for duration $T$. This expansion can be written as 
\begin{equation}
\theta_n(T) = \frac{\gamma T}{2}sin(\vec{k}\cdot\vec{r}+\phi_n) \begin{bmatrix}
    G_x & G_y & G_z
\end{bmatrix}
\cdot
\begin{bmatrix}
\xi_x \\ \xi_y \\\xi_z
\end{bmatrix}
\label{eq:vectorSolution}
\end{equation}
Where $\gamma$ is the gyromagnetic ratio of hydrogen, $\vec{k}$ is the acoustic wave number, and $\phi_n$
 is a difference in phase between the oscillating gradient field and the acoustic wave chosen by the operator. 
Equation (\ref{eq:vectorSolution}) presents a more accurate representation of observed voxel phase for cases where the MRH encoding gradient field varies spatially, as in the case of the human compatible MRH design~\cite{MRH2025}. Total observed phase, therefore, depends on the relative parallel orientation of the displacement and gradient vectors. The MRbOH method attempts to remove this directional dependence by resolving the individual components of $\hat{\xi}$ before calculating pressure.

Equation (\ref{eq:vectorSolution}) suggests that the orthogonal components of $\vec{\xi}$ at location $r$ can be obtained by observing a series of $\theta_n(T)$ as the gradient field $\vec{G}$ varies in orientation across $j\in \{0,\dots, J-1\}$ measurements, producing the matrix equation
\begin{equation}
\vec{\theta_n}={\Lambda_n}\mathbf{G}\vec{\xi}
   \label{eq:MRbOhMatrixEquation} 
\end{equation}
 where 
 \begin{equation}
 \vec{\theta_n}=
 \begin{bmatrix}
     \theta_{1,n}, 
     ...
     \theta_{J,n}  
 \end{bmatrix}^T
     \label{eq:BigTheta}
 \end{equation}
 
  \begin{equation}
\Lambda_n = sin(\vec{k}\cdot\vec{r}+\phi_n)
     \label{eq:kDefinition}
 \end{equation}
 and 
 \begin{equation}
 \mathbf{G}= \frac{\gamma T}{2}
 \begin{bmatrix}
     G_{x,1} & G_{y,1}& G_{z,1} \\
     \vdots  & \vdots & \vdots \\
   G_{x,J} & G_{y,J}& G_{z,J} \\ \\
 \end{bmatrix}.
     \label{eq:BigG}
 \end{equation}

Note that the arbitrary phase delay, $\phi_n$, and it's subscript $n$, is independent of any choice of $\vec{G_{k,j}}$.

When $\mathbf{G}$ is not singular, $\vec{\xi}$ can be resolved by computing $\mathbf{G}^{-1}\vec{\theta_n}$. Therefore, implementing omni-directional sensitivity for a magnetic resonance hydrophone requires devising a coil design and acquisition strategy that yields a well-conditioned matrix $\mathbf{G}$ within a region of measurement. 

\subsection{The Nuisance Factor, $\Lambda_n$} \label{sec:Lambda}
The term $\Lambda_n$ in equation \ref{eq:MRbOhMatrixEquation} is a nuisance factor that impedes resolving $\vec{\xi}$ by amplifying noise upon inversion. Both Passe-Carlus ~\cite{MRH2025} and Evans~\cite{Evans2004} accounted for $\Lambda_n$ by  limiting pressure estimation to MR voxels where $\Lambda_n$ approximated $\pm1$. While this method was able to produce results, it remained susceptible to noise.

A more sophisticated approach can be borrowed from MR elastography\cite{Kramer2025} where one acquires repeated measurements but with incrementing phase such that. 
\begin{equation}
\phi_n =2\pi \frac{n}{N}, n \in \{0,\dots, N-1\}
    \label{eq:phaseSteps}
\end{equation}
After inverting (\ref{eq:MRbOhMatrixEquation}), each component, $\xi_k$, can be estimated by evaluating, at frequency $\frac{1}{N}$, the magnitude of the discrete Fourier transform and multiplying the result by a factor of $\frac{2}{\sqrt{N}}$:
\begin{equation}
\xi_k = \frac{2}{\sqrt{N}}\Big|\frac{1}{\sqrt{N}}\sum _{n=0}^N \xi_{k}\lambda_n e^{-j2\pi \frac{n}{N}} \Big| 
    \label{eq:noiselessEstimate}
\end{equation}
Finally, under the plane wave assumption, estimated scalar acoustic pressure, $\hat{p}$ can then be obtained from $\vec{\xi}$  as follows: 
\begin{equation}
\hat{p} = \rho c \omega\sqrt{\sum_k^K \xi_k^2}
    \label{eq:pressureEstimate}
\end{equation}
where $\rho$ is the material density, $c$ is the sound speed, $\omega$ is the angular frequency of the wave, and $k$  indexes through the $K$ dimensions of $\vec{\xi}$.  Note that this technique now requires $JN$ acquisitions. All calculations must be repeated for each voxel location in the measurement field. 

\subsection{Noise Model} \label{sec:NoiseModel}
This section develops a statistical noise model for the final pressure estimate $\hat{p}$ . MR voxels are subject to additive, zero mean, uncorrelated, complex valued, Gaussian noise with variance $\sigma^2$ ~\cite{Gudbjartsson1995}.  In the frequently observed case where $\frac{A}{\sigma}>>3$ , where $A$ is voxel magnitude, voxel phase also follows zero-mean Gaussian, additive noise with variance $\frac{\sigma^2}{A^2}$.  

Under these assumptions, noise in $\xi_k$, after Fourier transform, but before the absolute value is taken, is also complex valued, zero mean, Gaussian, additive, and uncorrelated with variance of 
\begin{equation}
    \sigma^2_{\hat{\xi},k} = \frac{2\sigma^2 }{NA^2 s_1^2} (\mathbf{V\Sigma^{-2} V^*})_{kk}
    \label{eq:postFTVariance}
\end{equation}
Where we have substituted the matrix $\mathbf{G}$ with its singular value decomposition $\mathbf{G}=s_1 \mathbf{U \Sigma V^*}$
and 
\begin{equation}
    \mathbf{\Sigma}=\frac{\gamma T}{2}\mathrm{diag}(1, \frac{s_2}{s_1}, \frac{s_3}{s_1}) 
\end{equation}

Equation (\ref{eq:postFTVariance}) describes important features about the MRbOH technique. First, the singular values of $\mathbf{G}$ represent the sensitivity of the technique to ultrasound propagating along orthogonal axes. Second, as reported by Passe-Carlus et al~\cite{MRH2025}, the magnitude of $\mathbf{G}$ approaches zero with distance from the MRH electromagnet. This reduction decreases $s_1$ and therefore amplifies the observed noise. The factor $s_1$, therefore, can act as an objective against which MRH coil design can be optimized. Third, the condition number of $\mathbf{G}$, $\kappa=\frac{s_1}{s_3}$, represents worst-case noise amplification and presents a second objective against which MRbOH design can be optimized. 

We also note that, since the real and imaginary channels of \ref{eq:noiselessEstimate} produce uncorrelated noise with equal variance, the the absolute value operator transforms the bi-variate complex gaussian, noise into a Rice distribution~\cite{rice1944mathematical}. 

\subsection{Rectified Noise in MRbOH Pressure Estimates} \label{sec:RectifiedNoise}
The propagation of noise into (\ref{eq:pressureEstimate}) can be complicated by correlations along the $k$ dimension.  The degree of correlation varies with the construction of $\mathbf{G}$. However, this correlation decays as the condition number $\kappa$  approaches 1. As will be shown below, this study confines analysis to regions where $\kappa \approx 1$. We therefore assume with some confidence that noise in $\xi_k$ are uncorrelated, identically distributed Gaussian variables with a mean of zero and variance of $\sigma^2_\xi =\frac{2\sigma^2 }{NA^2 s_1^2}$. Under these conditions, the observed pressure, $\hat{p}$, follows a non-central $\chi$ distribution with $2K$ degrees of freedom and non-centrality parameter $\lambda_\chi=\frac{|\vec{\xi}|}{\sigma_\xi}$. 
The mean, $\mu_{\hat{p}}$, and variance, $\sigma^2_{\hat{p}}$, of $\hat{p}$ can then be written as
\begin{equation}
   \mu_{\hat{p}}= \sigma_{\xi}\rho c \omega \sqrt{\frac{\pi}{2}}L^{K-1}_{1/2}(\frac{-\lambda_\chi^2}{2})
   \label{eq:ChiMean}
\end{equation}
\begin{equation}
   \sigma^2_{\hat{p}} = \big(\sigma_\xi \rho c \omega \big)^2 \big(2K+\lambda_\chi^2  \big)-\mu_{\hat{p}}^2
   \label{eq:ChiVariance}
\end{equation}
Where $L^{K-1}_{1/2}$ represents the generalized Laguerre function~\cite{miller1964multidimensional, johnson1994continuous}. Note that, in the case of $K=1$, the noise reverts to a Rice distribution, as discussed above. The mean and variance of $\hat{p}$ under worst-case conditions can be estimated by substituting $\sigma^2_\xi =\frac{2\sigma^2 \kappa^2}{NA^2 s_K^2}$.

The most critical property of (\ref{eq:ChiMean}) is that when $\lambda_\chi$ grows small, $\mu_{\hat{p}}$ becomes larger than the true pressure, $p$. As discussed above, for human compatible MRH coil designs, $s_1$ decays quickly with distance. Thus, regions in the resulting MRbOH measurement are nearly guaranteed to over-estimate pressure. Overestimation of noisy data is a common factor for rectified, noisy signals and many references, including Constantinides et al.~\cite{Constantinides1997} and Gubjartson and Patz ~\cite{Gudbjartsson1995}, discuss techniques to remove bias in the presence of non-central, $\chi$ distributed noise. 

A second important property of (\ref{eq:ChiMean}) is that the amplification factor produced by noise can be, to some degree, anticipated from a small number of readily obtained parameters. In all published MRH experiments, the terms $\mathbf{G}$, $\rho$,  $c$, and $\omega$ are assumed to be known with high confidence~\cite{MRH2025,Plewes2000,Walker1998}. The singular values $s_k$, can be obtained from prior knowledge of the MRH encoding gradients. Measures of $A$ and $\sigma$ can be easily estimated from magnitude MR images. Thus, the majority of the terms in \ref{eq:ChiMean} can be obtained with high confidence and used toward MRbOH design. 

The following sections describe the design, construction, and validation of a MRbOH device using the design criteria proposed in this current section.  Due to constraints on equipment stability, this study limits it's experimentation and analysis to the 2D case where $K=2$.

\section{Methods}
The MRbOH concept was tested using a custom built apparatus that simultaneously permitted multiple MRH encoding coil locations as well as ultrasound transducer propagation angles. The apparatus was then scanned in an MRI system under many permutations of both coil location and transducer angle. The resulting MR images were processed to produce estimates of acoustic pressure as a function of propagation angle that were subsequently compared to estimates produced by calibrated hydrophones.  These methods are described as follows. 

\subsection{Transducer Design, Construction, and Characterization} \label{sec:Txducer}
\subsubsection{Design and Fabrication} \label{TxDesign}
The MRbOH technique was tested using a custom-fabricated, unfocused, single element transducer comprising a a 30 mm diameter, 4.2 mm thick piezo disk (SM411, STEINER \& MARTINS, INC., Davenport, FL) directly soldered to coaxial  cabling (MRG5801, Belden, St. Louis, MO). The transducer shell consisted of a front housing for the active element and a back lid, both fabricated using clear stereolithography resin (Formlabs Clear, Formlabs Inc., Somerville, MA). The shell contained an MR-visible cavity to indicate a plane transecting the disk diameter in MR images. The active element was bonded into the front housing using a 1.3 mm thick, approximately 1.5:1 by weight mixture of aluminum oxide (12000 Grit Aluminum Oxide, GANGOU) and epoxy (EA M-31CL, Loctite, Westlake, OH). The element was backed with a low density epoxy mixture made by mixing increasing amounts of glass microspheres (Fasco Epoxies Inc., Fort Pierce, FL) into epoxy until formation of a thick paste. The back lid was then applied and the transducer was allowed to cure for 24 hours before use. The resulting transducer had an electrical impedance of 150.6 $-$ $j$57.9 $\Omega$ at 500 kHz. The transducer was driven by a function generator (SDG2042X, Siglent, Solon, OH) and amplifier (240 L, E\&I, Rochester, New York) with no intervening matching network.

\subsubsection{Transducer Characterization} \label{sec:TxducerCharacterization}
The transducer was mounted on a plastic frame and submerged in a tank of deionized, degassed water. Two-dimensional pressure fields in a plane transecting the diameter of the transducer element were acquired using calibrated hydrophones connected to a motorized triaxial stage. This measurement was repeated over a total of five separate needle hydrophones with differing sensitivities, diameters, and manufacturers as reported in Table \ref{tab:hydrophones}. The table additionally reports estimated effective hydrophone diameter and the associated correction factor calculated via the methods of Wear \cite{Wear2026}.

\begin{table}
\caption{Hydrophone Models Used in This Study}
\label{tab:hydrophones}

\setlength{\textwidth}{3pt}

 \begin{tabular}{p{60pt}p{60pt}p{60pt}p{60pt}p{60pt}p{60pt}}
 \hline\hline
Model&  Manufacturer&  Diameter (mm)&  Uncertainty& Effective Diameter (mm)& Correction Factor\\
\hline
HNR-1000&  ONDA&  1&  $\pm1.5$ dB& 1.719& 1.0052\\
HNP-0400&  ONDA&  0.4&  $\pm1.5$ dB& 1.2149& 1.0026\\
NH0200&  PA&  0.2&  $18\%$& 1.0496& 1.0019\\
NH1000&  PA&  1&  $18\%$& 1.719& 1.0052\\
NH0500&  PA&  0.5&  $18\%$& 1.2981& 1.0030\\ 
\hline\hline
 \multicolumn{6}{p{180pt}}{PA: Precision Acoustics}
\end{tabular}
\end{table}

During measurement, each hydrophone was aligned with the center axis of the transducer using a removable fiducial and then rastered in 0.5 mm steps through a 40 mm wide by 63.5 mm long measurement plane. During measurement, the transducer was pulsed with a 50 cycle, 200 Vpk-pk sinusoid. Hydrophone signals at each step were measured using a digital oscilloscope (SDS2104X, Siglent, Solon, OH). The hydrophone signal was windowed to encompass 20 complete steady state cycles. Then, a fast Fourier transform was used to find the signal amplitude at 500 kHz. This amplitude was then converted to pressure using the hydrophone's calibrated sensitivity value. After all hydrophone scans had been completed, spatial averaging correction values were evaluated and applied via \cite{Wear2026} using the apparent beam full-width at half max (FWHM) observed by the NH0200 hydrophone. A final pressure estimate, and it's associated error, was computed from the estimates provided by each hydrophone using the weighted mean method described by Robinson et al. ~\cite{Robinson2006} and the manufacturer-provided worst-case estimate of each hydrophone's uncertainty. Estimated peak rarefactional pressure in the MR imaging region, assuming wave linearity, was 522 kPa, which corresponds to a non-derated mechanical index of 0.74 which is within biophysical safety recommendations ~\cite{Aubry2025_itrusstSafety},~\cite{MARTIN2024607}.

\subsection{MRbOH Coil Design and Characterization} \label{MRH Coil}

\begin{figure}[H]
\centerline{\includegraphics[width=300pt]{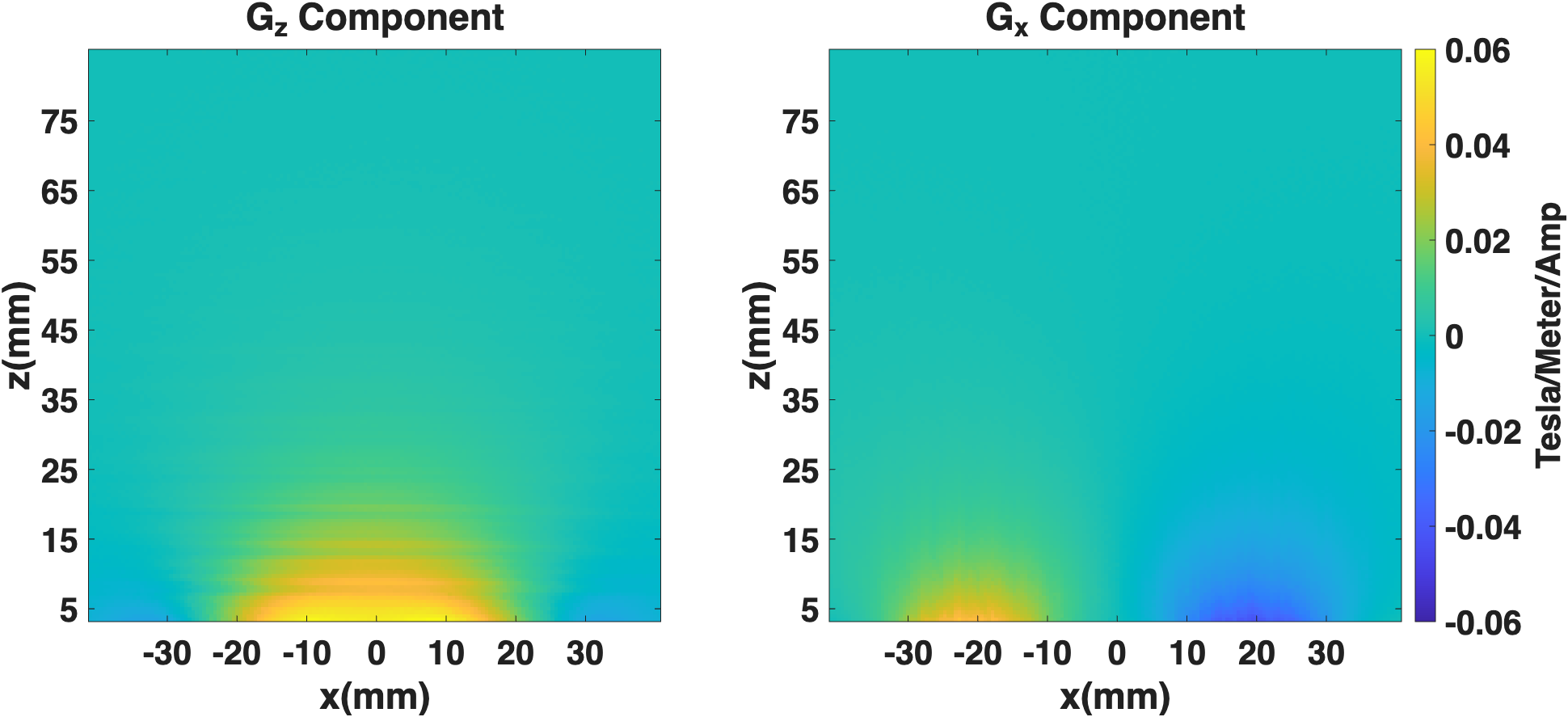}}
\caption{Magnetic field gradients produced by the ultrasound sensitizing coil used in this study. The gradients produce spatially varying sensitivity to orthogonal propagation directions, with peak sensitivity along each axis occurring at separate locations relative to the coil center. }
\label{fig:gradFields}
\end{figure}

\subsubsection{MRH Coil Design and Driving Method} \label{sec:MRH Design}
The sensitizing MRH coil mimicked that described by Passe et al. ~~\cite{MRH2025}. The design maximizes the first singular value, $s_1$, at a depth of 20 mm along the $z$ axis from the coil surface. The coil's magnetic field gradient was characterized using the methods described by Passe et al ~\cite{MRH2025}.  Briefly, a Teslameter (F71, Lake Shore Cryotronics, Westerville, OH) was rastered through a 158 mm wide by 78 mm deep plane offset 4 mm from the coil face using the same step sizes as used to characterize the hydrophone field described above. The coil was energized by 5 A of constant current and the Teslameter recorded the component of the magnetic field that was anticipated be parallel to the MRI scanner bore. The magnetic field gradient for each orthogonal axis in the image was then calculated by a 3rd order spatial finite differencing scheme and trimmed to equal size.

\subsubsection{MRH Placement Optimization} \label{sec:MRH Placement}
The MRbOH acquisition scheme used in this study was inspired by Figure \ref{fig:gradFields}, which displays the $G_x$ and $G_z$ components of the encoding field gradient and corresponds to MRH sensitivity along the \textit{z} and \textit{x} axes, respectively, where the $z$ axis is parallel to the MRI bore. In the figure, maximum sensitivity to acoustic displacement along the \textit{z} and \textit{x} directions occur in physically different locations. A well-conditioned matrix $\mathbf{G}$ might be constructed for a measurement region placed in front of the coil by taking additional MRH measurements in the same physical location but with the MRH coil displaced along the $\pm x$ direction. 

A suitable coil displacement value was estimated in silico. The coil geometry described above was replicated in a magnetostatic finite-element software package (COMSOL, Burlington MA) with the center of the coil surface placed at the origin and the coil's axis of symmetry along the \textit{z} direction. The magnetic field gradients $G_z$ and $G_x$ produced by the coil were retained within a 2D measurement region centered at the origin in the \textit{z} direction and 30 mm distant along \textit{x} from the coil face. Gradient field measurements were then recomputed but with the coil translated in the  $\pm$\textit{x}. directions by varying amounts. For each vertex location in the simulation, gradient measurements were used to construct a $3 X 2$ matrix $\mathbf{G}$ and then calculate the matrix's condition number, $\kappa$. A suitable coil displacement value was then obtained by choosing the displacement that minimized the average condition number within a 30 mm circular region of interest, resulting in an optimal displacement of 30 mm. This displacement was  used in the experiments described in section \ref{sec:permute}.

\subsection{Experimental Apparatus}  \label{sec:holder}
An experimental apparatus to test the MRbOH mechanism was designed and 3D printed out of PLA with total dimensions of 208 x 144 mm and internal tank dimensions of 109 x 98 mm. When placed in an MRI scanner, the apparatus aligned the MRH coil axis with the scanner's bore axis, which is labeled the $z$ axis. It included a rail system that supported the translation of the coil along the $x$ axis, with locking mechanisms placed at 10 mm intervals. The PLA structure was waterproofed with an epoxy coating and then lined with the custom acoustic absorber and scattering material described below.  Inside the structure resided supports for a gelatin brick.

\begin{figure}[H]
\centerline{\includegraphics[width=300pt]{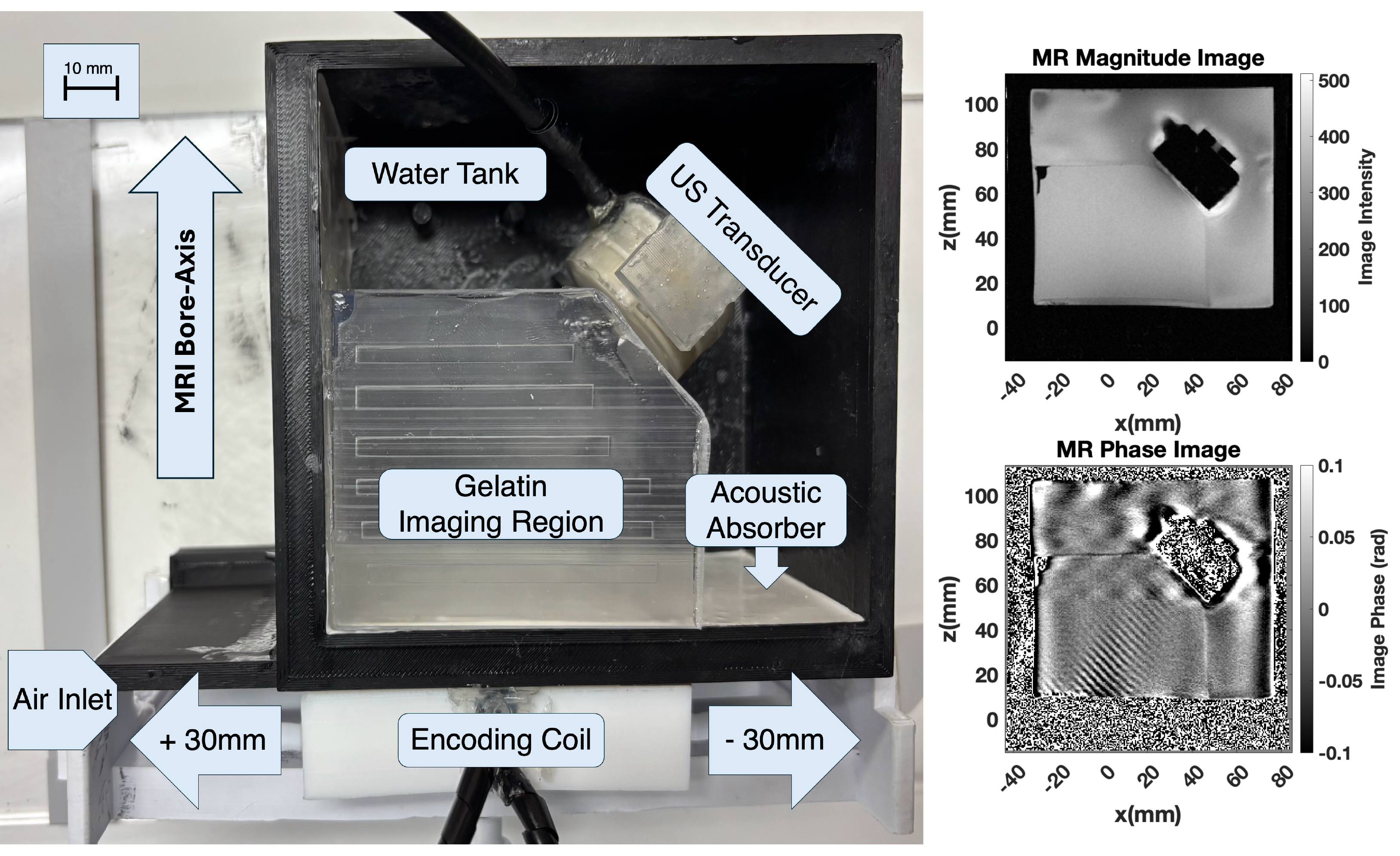}}
\caption{Left: Experimental apparatus used to test the MRbOH technique. The water tank is empty for photographic purposes. The transducer is shown aligned with one of three possible faces. And the encoding coil is placed in the central location. One of two acoustic absorbers are visible in the photograph. Right: Example MR image magnitude and phase images displaying, respectively, visible transducer and wavefront contrast.}
\label{fig:apparatus}
\end{figure}

\subsubsection{Transducer Orientation}
The apparatus also contained locking mechanisms that held the transducer at angles of $0^o$, $45^o$, and $90^o$ relative to the $z$ axis. They  were designed such that the same area of the transducer's acoustic far field, no matter the propagation angle, propagated through a common circular region 30 mm distant along the $z$ axis from the origin. Additionally, the apparatus was designed so that a plane transecting the diameter of the transducer at any location also transected the diameter of the MRH encoding coil. This plane coincided with the planes acquired by the Teslameter and hydrophones described in sections \ref{sec:TxducerCharacterization} and \ref{sec:MRH Design}.
\subsubsection{Gelatin Phantom}
The gelatin brick phantom (7\% w/w , 250 Bloom , Vyse Gelatin, Morrisville, NC).) was shaped to produce 3 flat faces against which the transducer could be placed. Gelatin density was measured directly by making a disk of known volume and measuring its weight using a digital scale. Gelatin sound speed was measured using a pulse-echo method at 500 kHz~\cite{Zeqiri2010} with the speed of sound in water corrected to the temperature of the water tank~\cite{lubbers1998simple}. Average gel pressure attenuation was found to be less than .05 dB cm$^{-1}$ in a through-transmit measurement.
\subsubsection{Reflection Reduction} \label{sec:Absorber}
To minimize reflections, we developed acoustic absorbers using a two-layer design. The front layer consisted of a 1:1.12 w/w mixture of silicone and aluminum oxide cast into a rapid prototyped mold which was patterned with a grid of 3.76 x 3.76 x 7 mm pyramid structures. After curing, the mold was removed and the resulting space was filled with an  approximately 1:1.12:0.08 by weight mixture of silicone, aluminum oxide, and glass microspheres, forming a 10 mm thick, two component absorber. Pulse-echo measurements at 500 kHz demonstrated at least 80 percent reduction in reflected pressure compared reflections off an aluminum plate. These absorbers were integrated into the two sides of the 3D-printed structure that were directly impinged by the acoustic field.

\subsubsection{Acoustic Coupling}
Space not occupied by the absorbing material, transducer, or gelatin was filled with degassed, de-ionized water in order to provide acoustic coupling. During experiments, the coil was air cooled using a compressor (VFC209P-5T, Fuji Electric Co., Tokyo, Japan) and plastic hose fed into the room. The apparatus is displayed in Figure \ref{fig:apparatus}. The figure also displays example MR magnitude and phase images. The transducer, gel, and water bath are apparent in the magnitude images. The phase image displays sinusoidal phase encoded through the MRbOH technique.

\subsection{MRH Scan Sequence} \label{sec:PulseSequence}
The apparatus was placed at iso-center in a commercial, 3T, MRI scanner (Vida, Siemens, Erlangen). The scan plane was prescribed to transect both the fiducial marker in the transducer housing and a fish oil capsule, when observed, placed in the center of the encoding coil.  The apparatus then had a 330 mm diameter acrylic half-cylinder placed over it and was encompassed by an 8-channel spine array coil below and by an 18 channel abdominal coil above. MRH scans of the plan transecting the fiducial markers were completed using the spin-echo sequence reported by Passe-Carlus et al. ~\cite{MRH2025}. using a TR of 1 s for all acquisitions, a resolution of 0.5 x 0.5 mm with a slice thickness of 5 mm., a scan field of view of 128 mm by 128 mm. The total acquisition time for a single image was 256 s. 

\subsection{MRbOH Pressure Characterization}

\begin{figure}[H]
\centerline{\includegraphics[width=300pt]{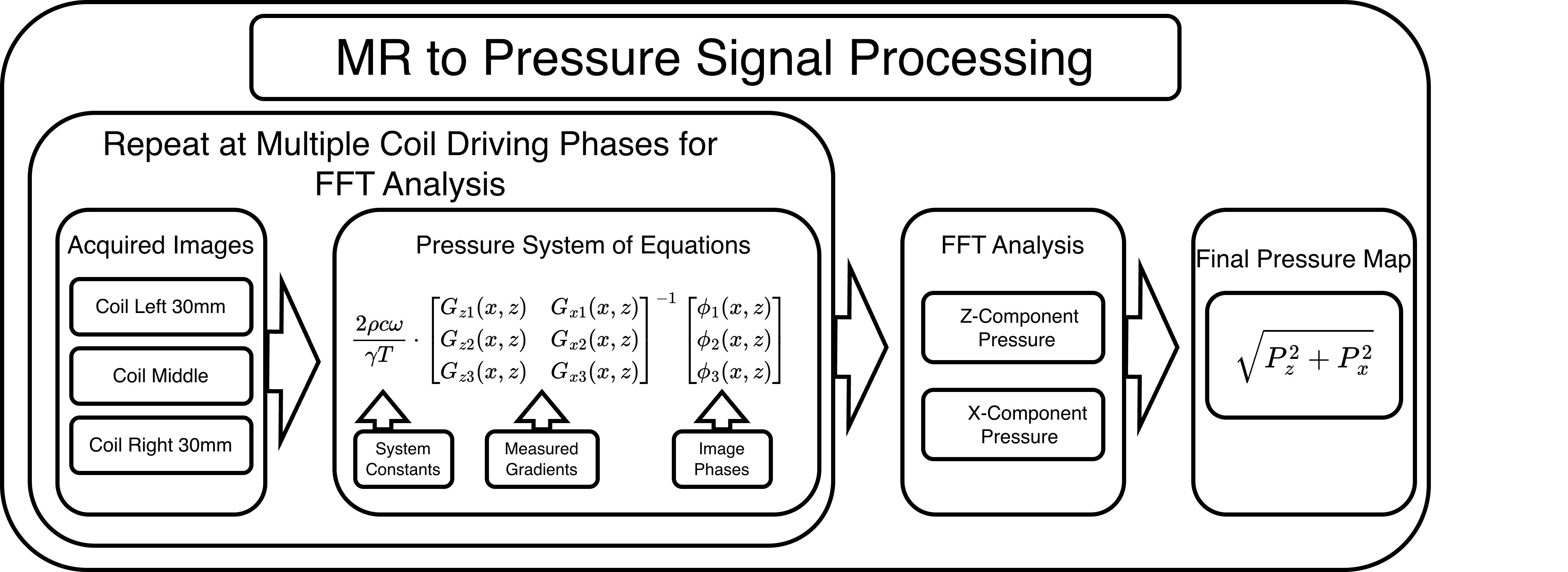}}
\caption{Schematic of the processing pipeline used to estimate acoustic pressure in all experiments.}
\label{fig:flowchart}
\end{figure}

\subsubsection{Experimental Permutations} \label{sec:permute}
To test the MRbOH technique, the MRH scan sequence was repeated under a nested loop strategy. For each of the three transducer propagation angles, the scan sequence was repeated with the MRH encoding coil placed at -30, 0, and +30 mm along the $x$ axis, setting $J=3$.  For each gradient coil location, the scan sequence was repeated as the phase difference between the gradient and ultrasound, $\phi_n$,  was incremented, with $N=4$. Thus, for each transducer location, a MRbOH measurement required $J*N=12$ scans, for a total of approximately 51 minutes of scanning per transducer angle. For these experiments, the TE of the scan was set to 115 ms, for a total encoding time of 93.2 ms, split equally by the sequence's spin echo RF pulse.  A 4 minute break between each scan was imposed to allow the MRH coil and amplifier to cool, creating a total experimental time of 297 minutes.

\subsubsection{Transducer Operation} \label{sec:TxVoltage}
When triggered by the MRI pulse sequence, as described by Passe et al.~~\cite{MRH2025}, the transducer was driven by a 200 Vpk-pk 500 kHz sinusoidal CW wave over each 46.6 ms burst allowed by the MRH pulse sequence. As described by Plewes and Passe, the phase of every other burst was rotated by $180^o$ ~\cite{Walker1998, MRH2025} to prohibit the spin echo effect from rewinding ultrasound-induced phase. Voltages across the transducer were monitored and varied $\pm$ 3\% throughout the entire experiment.

\subsubsection{MRH Coil Operation} \label{sec:CoilCurrent}
The MRH coil was driven with a 500 kHz sinusoidal continuous wave during the same periods as the transducer. The peak current varied with coil and amplifier temperature and was monitored continuously using a current probe (CWT UM/1/B, GMW Associates, San Carlos, Ca). Average coil current for all acquisitions was $\approx$52 Amps pk-pk. The peak current observed during the last TR of each scan was recorded and then used, in conjunction with the coil characterization provided by section \ref{sec:MRH Design}, to estimate the magnetic field gradient vector, $\vec{G_o}$, at a given voxel location.

\subsection{MRboH Noise Characterization} \label{sec:NoiseMethod}
The noise characteristics of the MRbOH technique were examined by repeating the pressure characterization experiment but with the ultrasound transducer deactivated and removed from the apparatus during the experiment. With the transducer removed, phase shifts, $\theta_n$, and transducer propagation angles produce no effect on image quality. Therefore, for each of the three gradient locations, the MRI pulse sequence simply acquired five images, producing a total of 15 images. Additionally the TE was shortened to 60 ms, for a total MRH encoding time of 41.6 ms.  Total experiment time was 121 minutes.

\subsection{Data Processing and Analysis}
\subsubsection{Region of Interest for Analysis}  \label{sec:ROI}
All measurements were imported into MATLAB (R2024a, MathWorks, Natick, MA, USA) for analysis.  Quantitative analyses were conducted on a 15 mm diameter circular region of interest placed 50 mm from the face of the transducer. As described in section \ref{sec:holder}, this region resided at the same MR image coordinates no matter the transducer location. This region was identified in magnitude MR images by measuring 50 mm from the center of the outline of strong image contrast produced at the transducer-gel boundary. The width of this region was equivalent to the FWHM of the transducer's expected field at 50 mm distance from the transducer surface. It also corresponded closely to the circular region used to optimize coil displacement spacing in section \ref{sec:MRH Placement}.

\subsubsection{Gradient Field Alignment} \label{sec:Alignment}
The magnetic field gradients measured by the Teslameter were co-registered into the MRI coordinate space using a manual method. The edge of the fiducial marker placed in the center of the coil was estimated in each magnitude MR image. The center of the Teslameter measurements obtained in section \ref{sec:MRH Design} were then aligned with this point and the Teslameter measurements were interpolated to the MR image coordinate space using nearest neighbor interpolation.

\subsubsection{MRbOH Pressure Calculation} \label{sec:PressureCalc}
For each transducer location, per-voxel pressure estimates were calculated from (\ref{eq:pressureEstimate}). More specifically, a $K=2$ version of the matrix $\mathbf{G}$ was constructed on a per-voxel basis within the region of interest using the aligned and interpolated Teslameter data. Then, on a per-voxel basis, the raw MR phase, was formed into the vector $\vec{\theta_n}$ and was used to invert (\ref{eq:MRbOhMatrixEquation}), followed by a calculation of  (\ref{eq:noiselessEstimate}) and then (\ref{eq:pressureEstimate}).  Figure \ref{fig:flowchart} presents a schematic of the processing pipeline. Finally, the beam FWHM and mean and standard deviation of pressure within the region of interest were calculated.

\subsubsection{MRbOH Noise Calculation} \label{sec:NoiseCalc}
 The precision of the MRbOH technique was assessed using a bootstrapping technique. The five scans, taken at each gradient location without the transducer active, were used to produce, in a combinatorial sense, at each voxel within the region of interest, a total of 125 separate vectors of observed voxel phase, $\vec{\theta_n}$.  These 125 vectors were then randomly sampled, with replacement, into 200 groups of four vectors in order to mimic the $N=4$ phase steps. Using the same methods as in section \ref{sec:PressureCalc}, these vectors were used to produce a total of 200 estimates of pressure at each voxel within the region of interest. The per-voxel mean and standard deviation within the region of analysis were then computed.

 \subsubsection{MRbOH Noise Prediction} \label{sec:NoisePrediction}
Magnitude image voxels within the region of interest for all 15 MR images taken without the transducer active were averaged together and then used to produce per-voxel regional estimates of $A$ as described in (\ref{eq:postFTVariance}). Meanwhile, $\sigma$ from the same equation was estimated by using the 15 magnitude images to create 15 difference images. The standard deviation of all pixels within the region of interest was computed, and then divided by $\sqrt{2}$ to compensate for the difference operator~\cite{NEMA2008}. These parameters were used to create a per-voxel prediction of the observed pressure mean and standard deviation within the region of analysis using (\ref{eq:ChiMean}) and  (\ref{eq:ChiVariance}). Because the transducer was deactivated during acquisition, the term $\lambda_\chi$ was known be $0$ with high precision.

\begin{figure}
\centerline{\includegraphics[width=300pt]{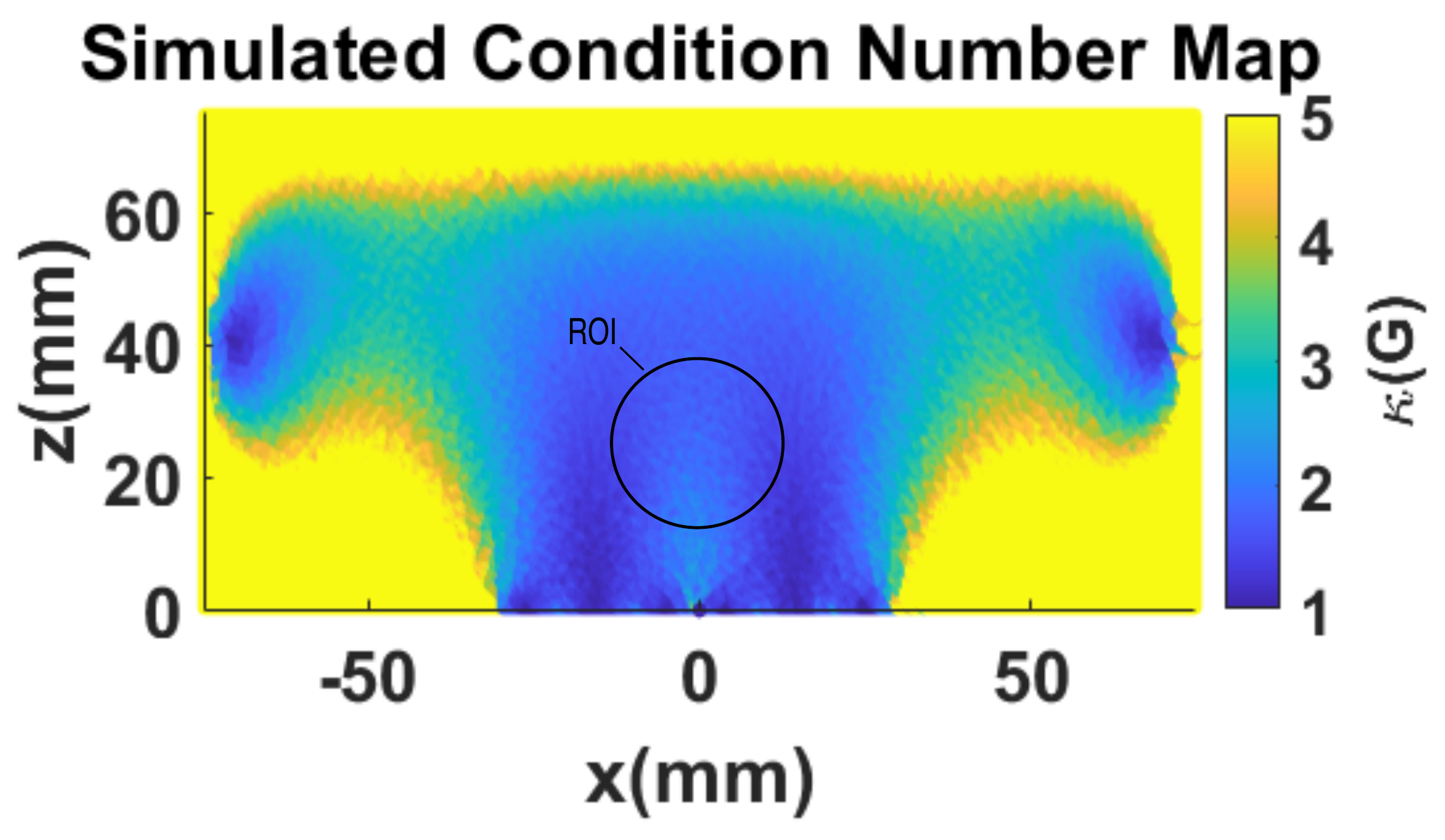}}
\caption{Condition number map produced by simulations of coil displacements of $\pm 30$mm.}
\label{fig:condNum}
\end{figure}

\section{Results}
The results of the experiments described in this study are reported as follows. 
\subsection{MRH Placement Optimization}
Figure~\ref{fig:condNum} displays a simulated spatial map of condition numbers for the $\pm$ 30 mm coil displacement value. Displacements at this value produce a large, irregularly shaped region containing a condition number below 5.  The average condition number within the region of interest marked in the figure is 1.8 and is the lowest average condition number produced by evaluated coil displacements. 

\begin{figure}[H]
\centerline{\includegraphics[width=300pt]{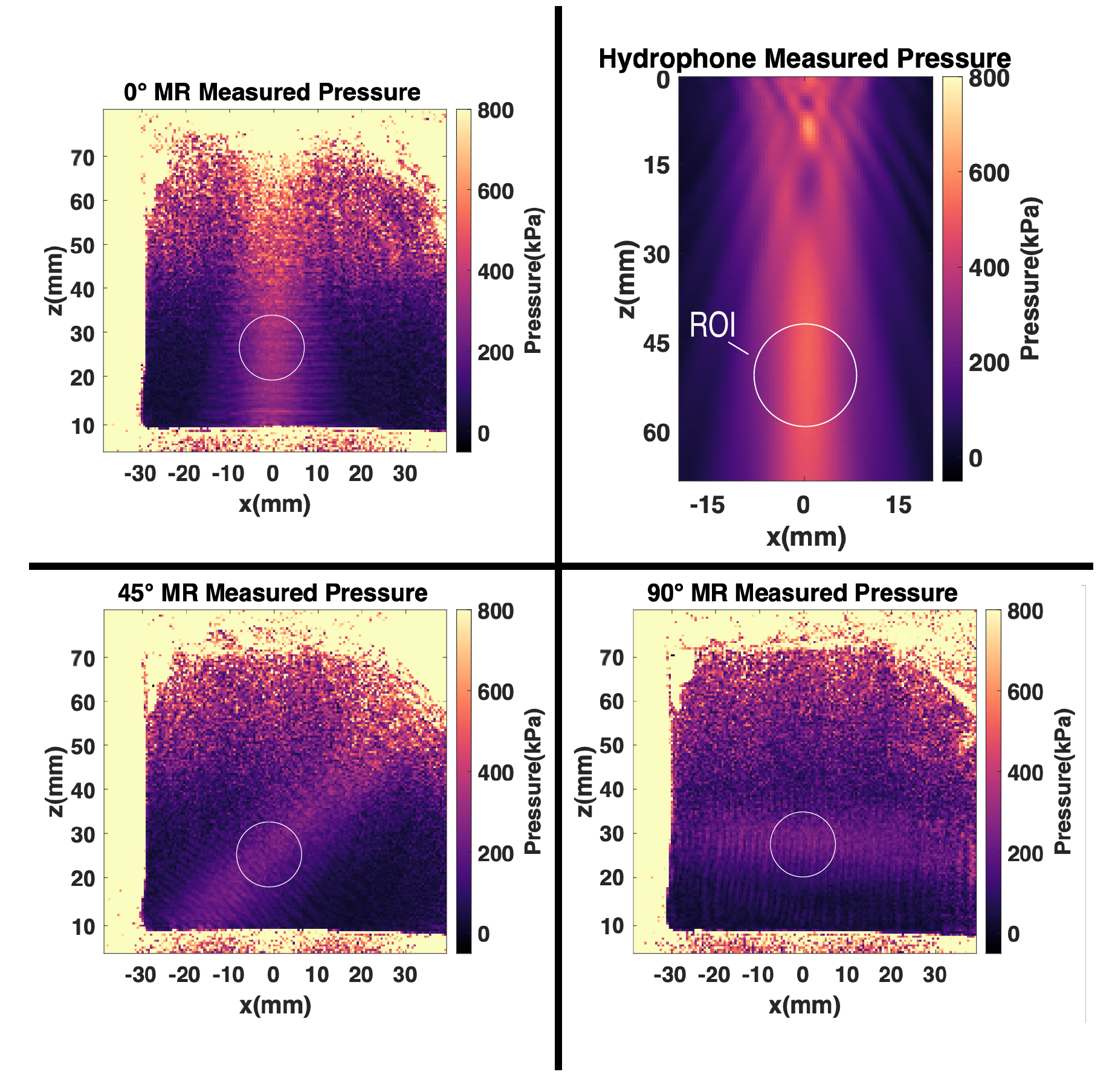}}
\caption{MRbOH and hydrophone maps of acoustic pressure. The region used for analysis is also indicated. The MRbOH method recapitulates the shape of the transducer's far field. Partial reflections are visible as $\lambda/2$ striations. Acquisitions at $45^o$ and $90^o$ display less sensitivity and more noise than the $0^o$ case.}
\label{fig:pressurePlots}
\end{figure}

\subsection{MRbOH Pressure Estimates}

Figure~\ref{fig:pressurePlots} displays maps of acoustic pressure produced by the MRbOH method for all three transducer placements along with the analysis ROI and the pressure field observed by the composite hydrophone. Regardless of transducer orientation, the MRbOH method is able to capture a slowly diverging field whose energy is mostly contained within a $\pm15$ mm width. Small ripples in pressure intensity occur at $1/2$ wavelength periodicity and represent standing waves produced by partial reflections. The sensitivity of the MRbOH method appears to be non-uniform, with MRbOH detecting lower magnitude pressure for the $45^o$ and $90^o$ cases.  Additionally, noise speckle in the $90^o$ case appears approximately 2X more apparent the $0^o$ case, which is consistent with a condition number, $\kappa$, of approximately 2. Additionally, noise obscures any MRbOH measurement of the transducer's complex near-field. 

\begin{figure}[h]
\centerline{\includegraphics[width=300pt]{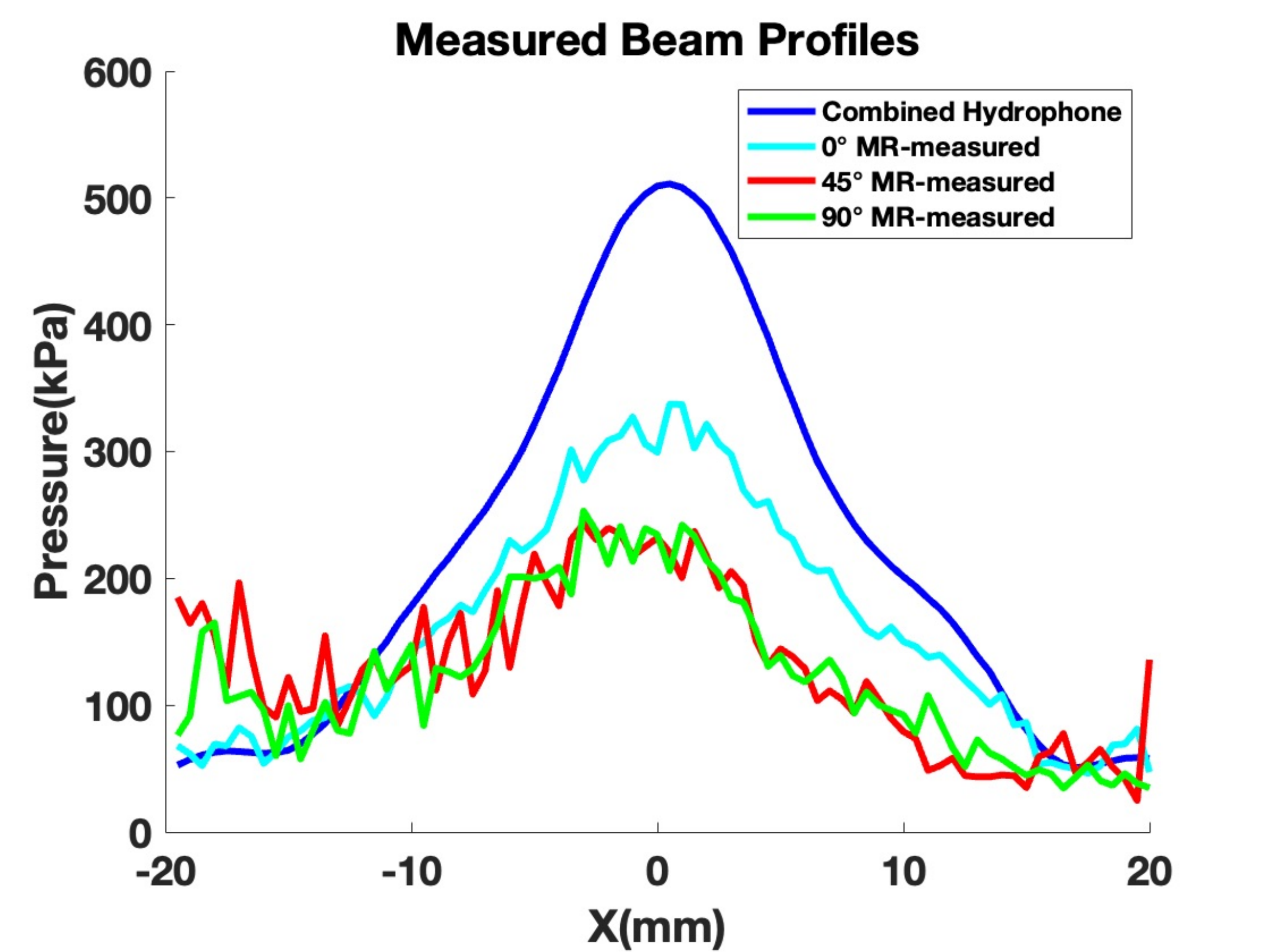}}
\caption{Averaged cross sectional view of the measured acoustic beam.}
\label{fig:profile}
\end{figure}

The ability of the MRbOH technique to capture lateral roll-off of the acoustic field can be seen in Figure \ref{fig:profile}, which plots cross sectional profiles of the pressure field returned from both the hydrophone and MRbOH. To reduce the effects of noise and partial reflections, the plotted MRbOH profiles are an average of three consecutive, individual profiles. These profiles are centrally taken 50 mm from the transducer face. Measured FWHM for the hydrophone profile was 14.5 mm. Measured FWHM for lowpass filtered MRbOH profiles were,  16, 15, and 17.5 mm for the $0^o$, $45^o$ and $90^o$ cases, respectively.

\begin{figure}[h]
\centerline{\includegraphics[width=300pt]{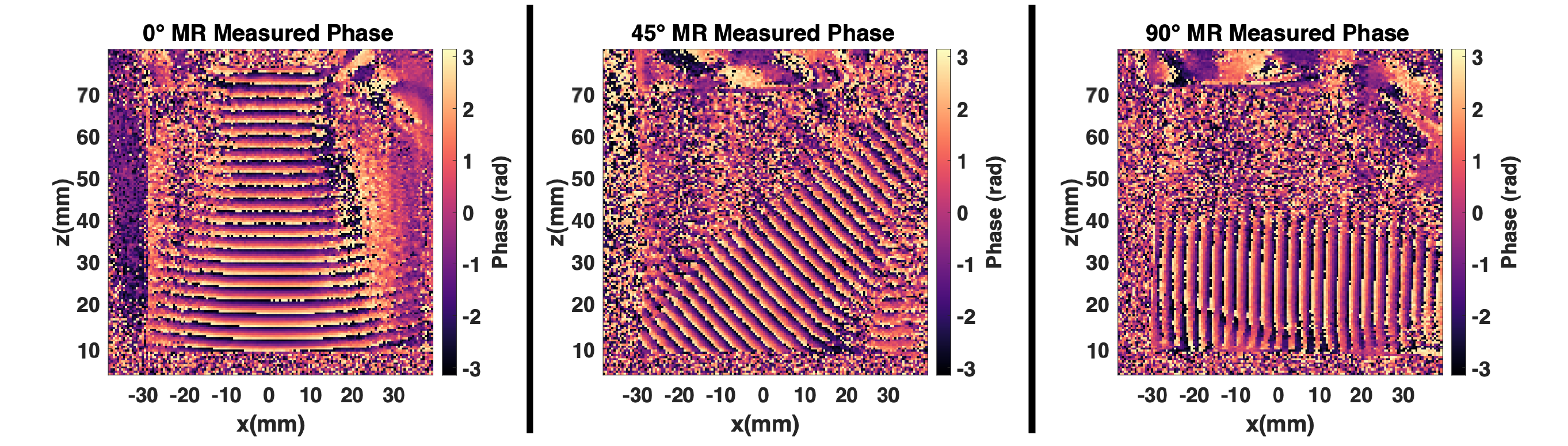}}
\caption{Complex phase produced by the MRbOH method representing relative phase of the acoustic wave. The phase patterns are consistent with a slowly diverging wavefront}
\label{fig:phase}
\end{figure}

Figure \ref{fig:phase} displays phase produced by replacing the absolute value operator with a complex-valued arc-tangent operator in~\ref{eq:noiselessEstimate}. Under a monochromatic plane wave approximation, the step size in phase wraps is equivalent to the beam's wavelength. The phase images display a wavelength of approximately 3 mm for all transducer orientations and corroborate with a slowly diverging beam as expected from the far-field of a circular planar transducer. Finally, this figure displays periodic wave fronts for regions that extend laterally well beyond the FWHM of the beam observed in the magnitude pressure images. The quality of the phase signal in regions with observed small pressure magnitude suggests that the lateral roll-off observed in Figs \ref{fig:pressurePlots} and \ref{fig:profile} is produced by roll off in the acoustic field and not a roll off in device sensitivity. 

\begin{figure}[H]
\centerline{\includegraphics[width=300pt]{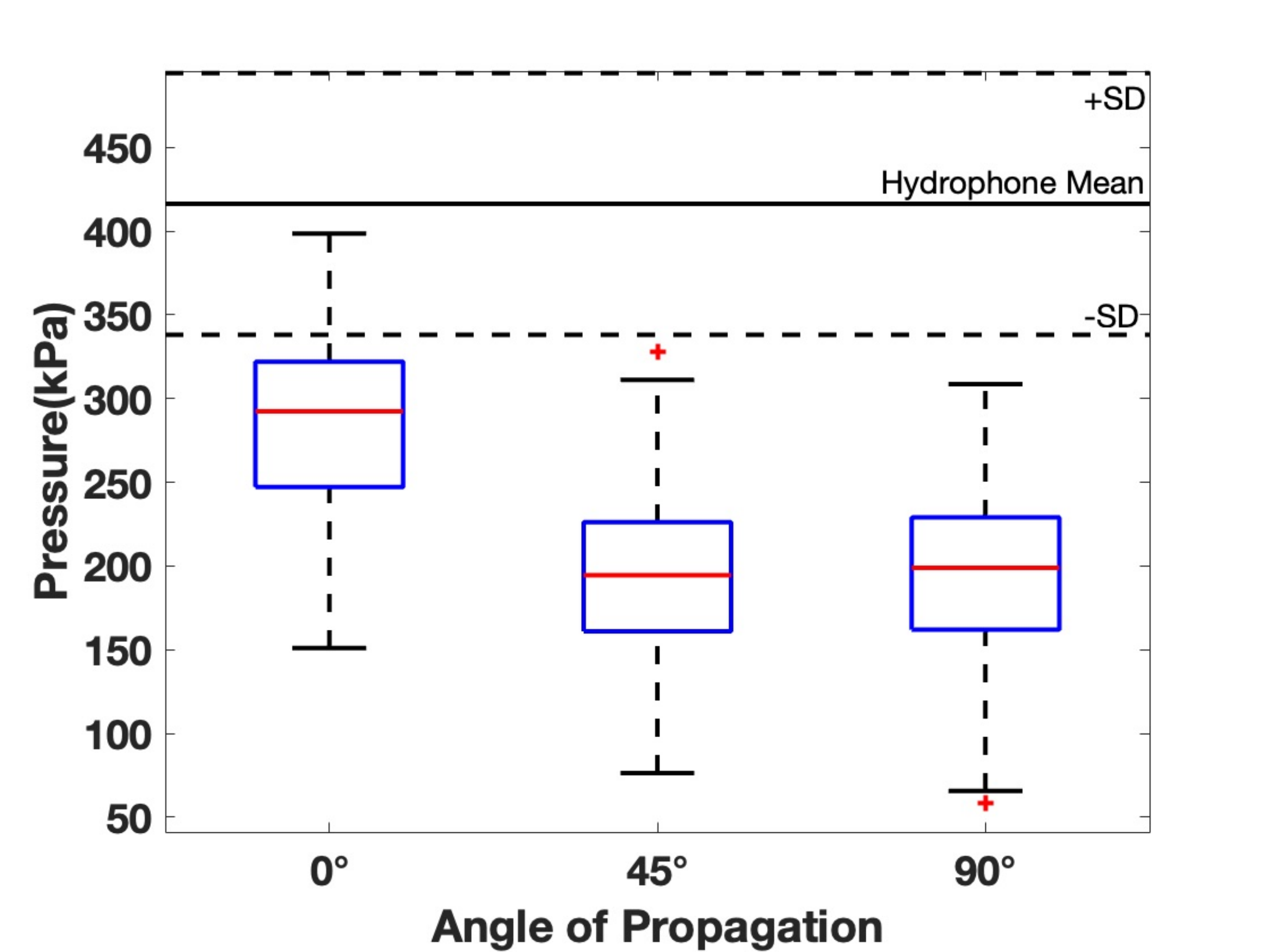}}
\caption{Box plots of per-pixel acoustic pressure observed by MRbOH within the region of interest as a function of ultrasound propagation angle compared to that recorded by the composite hydrophone.}
\label{fig:boxPlot}
\end{figure}

Figure \ref{fig:boxPlot} displays the median and inter-quartile range of pressures observed in the region of interest for all transducer angles as well as the mean and standard deviation of the pressures found by the composite hydrophone. The figure corroborates the evidence observed in Figure~\ref{fig:pressurePlots} that MRbOH underestimated pressure relative to the composite hydrophone and that this underestimation was more severe for the the $45^o$ and $90^o$ cases.  Curiously, the degree of underestimation does not appear to be linear with angle relative to the MRI bore. 

\subsection{MRbOH Precision and Noise Characteristics}

\begin{figure}[H]
\centerline{\includegraphics[width=\columnwidth]{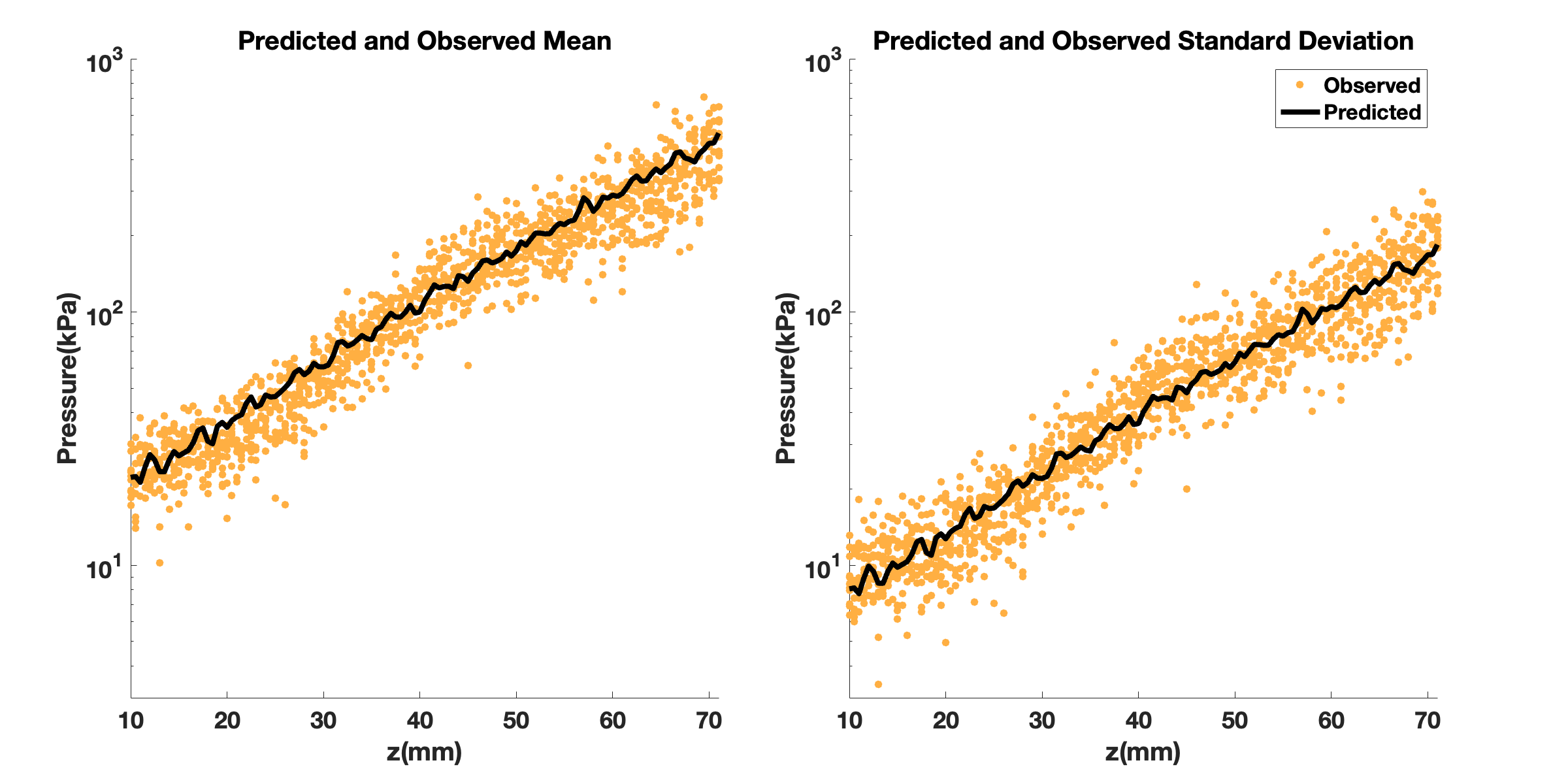}}
\caption{Plots showing the observed mean and standard deviation compared to the analytical prediction across a 5 mm wide central strip along the z axis. For the mean, $R^2 = 0.85$. For standard deviation, $R^2 = 0.83$}
\label{fig:noisePlot}
\end{figure}

Figure \ref{fig:noisePlot} plots the per-pixel mean and standard deviation of pressure observed when the transducer was deactivated.  These points are confined within a 5mm wide strip along the $x$ axis that extends from the encoding coil surface along the $z$ axis.  It also plots the predicted mean computed in section \ref{sec:NoisePrediction} above. The predicted mean and standard deviation display good agreement with observation with the coefficient of determination between observed and predicted values for the mean and standard deviation being $R^2 = 0.85$ and $R^2 = 0.83$ respectively.

\section{Discussion}
This study presents theoretical underpinnings and experimental validation of an MR based Omni-directional Hydrophone (MRbOH). Using a condition number as an optimization metric, this study was able to design and implement a MRbOH technique that detected relevant features of acoustic beams propagating at $0^o$, $45^o$, and $90^o$ relative to the MRI bore axis in a gel phantom. Figure. \ref{fig:pressurePlots} demonstrates evidence of standing waves as half wavelength spaced striations in the pressure magnitude. The estimated ultrasound phase profiles in \ref{fig:phase} agree with what would be expected from a slowly diverging planar wave. And beam widths found via the MRbOH technique match well with those found by calibrated hydrophones.

However, the MRbOH technique does not appear completely equivalent to traditional hydrophones. Within a co-registered circular region of interest, MRbOH measured pressures were 285.5 $\boldsymbol{\pm}$ 49.5 kPa, 193.13 $\boldsymbol{\pm}$ 43.55 kPa, 194.7 $\boldsymbol{\pm}$ 47.38 kPa at $0^o$ ,$45^o$ ,$90^o$ degrees of propagation, respectively. In contrast, pressure measured using needle hydrophones in the same region of was 416.19 $\boldsymbol{\pm}$ 78.07 kPa.  The MRbOH technique, therefore underestimated the hydrophone by 31.4\% , 53.6 and 53.2\% for $0^o$, $45^o$, $90^o$ cases, respectively.  These are dramatically larger degrees of underestimation than reported by Passe-Carlus et al. ($\approx12\%$)~\cite{MRH2025}   and Evans ($\approx0\%$)~\cite{Evans2004}.  One possible cause of this underestimation could be misalignment between the MR scan plane and the hydrophone and gradient measurement planes. There is likely at least small misalignment between these measurement planes because alignment was performed manually and the fiducial capsule placed in the center of the MRH coil was difficult to identify in some scans due to RF shielding by the coil windings. Other possible sources of error include motion artifacts induced by the blowing air which caused the coil to vibrate in place, standing waves within medium, and instabilities in equipment performance. The assumption of a monochromatic plane wave may also prove too faulty. Given the divergence in observed pressure error between this study, Passe-Carlus et al.~\cite{MRH2025}  and Evans~\cite{Evans2004}, the true accuracy of MR-hydrophone techniques remains uncertain. 

The study also presents a statistical signal model for the MRbOH technique. Figure. \ref{fig:noisePlot} demonstrate that rectified noise increases with depth and can overestimate pressure by at least tens of kPa. Additionally, (\ref{eq:ChiMean}) and (\ref{eq:ChiVariance}) accurately predict the observed mean and standard deviation across fluctuations in both gradient strength and image SNR. The good agreement between the observed and predicted measurement mean and standard deviations suggest that the noise model can inform both MRbOH design and signal processing.

This study contains several limitations. The MRbOH technique presented here assumes a single monochromatic plane wave and confines itself to a single, planar acoustic source and a homogeneous medium. However, the presence of standing waves in Figure \ref{fig:pressurePlots} demonstrates this assumption to be false. Further, the accuracy of the MRbOH technique in the presence of strongly focused beams, shocked beams, or complex media has yet to be addressed. The technique also assumes that acoustic impedance is known with high precision--which may not be true for pathologic tissue. Finally, the technique relies on long measurement times and yet also assumes that only the gradient field, $\vec{G}$, and phase delay, $\phi_n$  change throughout the measurement.  System instabilities, such as scanner drift, temperature fluctuations, and amplifier performance, as well as patient movement may all challenge this assumption. However, despite these limitations, the ability of the MRbOH technique to noninvasively capture acoustic fields propagating along arbitrary axes within biologic materials may still prove highly useful to TUSN applications, where skull acoustic properties exert strong influence over the treatment process.

\section{Conclusion}
MRH methods have previously been shown to be able to encode particle displacement under the limitation that the ultrasound must propagate along the MRI bore-axis. The MRbOH technique presented here expands upon previous studies to  non-invasively detect and quantify ultrasound waves without angular restriction, allowing for easier integration with clinically relevant treatment devices and geometries.

\section*{Acknowledgment}

The authors thank the BYU MRI Research Facility and the BYU Electrical and Computer Engineering IMMERSE Program for their generous support of this study. The authors declare that they have no conflicts of interest to disclose. his work was supported in part by the National Science Foundation under Grant 2138403, the National Institutes of
Health under grant R01EB032773 and the Focused Ultrasound Foundation Global Intern Program.

\bibliography{refs} 
\bibliographystyle{ieeetr}



\vfill
\end{document}